\documentclass[paper]{JHEP3}
\usepackage{epsfig,multicol}
\newcommand{\eqn}{\begin{eqnarray}}
\newcommand{\eqnx}{\end{eqnarray}}

\title{Semiclassical Geometry of 4D Reduced Supersymmetric 
Yang-Mills Integrals}

\author{Zdzislaw Burda\\
Institute of Physics, Jagellonian University\\
ul. Reymonta 4, 30-059 Krak\'{o}w, Poland\\
E-mail: \email{burda@th.if.uj.edu.pl}}
\author{Bengt Petersson\\
Fakult\"at f\"ur Physik, Universit\"at Bielefeld\\
P.O. Box 100131, D-33501 Bielefeld, Germany\\ 
E-mail: \email{bengt@physik.uni-bielefeld.de}}
\author{Marc Wattenberg\\
Fakult\"at f\"ur Physik, Universit\"at Bielefeld\\
P.O. Box 100131, D-33501 Bielefeld, Germany\\ 
E-mail: \email{wattenbg@physik.uni-bielefeld.de}}
\abstract{We investigate semiclassical properties of space-time geometry of the low energy limit of reduced four dimensional supersymmetric Yang-Mills integrals 
using Monte Carlo simulations. The limit is obtained
by a one-loop approximation of the original Yang-Mills
integrals leading to an effective model of branched polymers. 
We numerically determine the behaviour of the gyration radius, 
the two-point correlation function and the Polyakov-line operator 
in the effective model and discuss the results
in the context of the large-distance behaviour of
the original matrix model.
}

\begin{document}

\section{Introduction}
The ten dimensional matrix model of reduced
supersymmetric Yang-Mills integrals \cite{ikkt}
is believed to be a good candidate for a non-perturbative 
definition of string theory for the following reasons 
\cite{fkkt,aikktt,aikkt}.
In the large $N$ limit the action of the model takes
the form of the action of IIB strings.
The model is supposed to contain in the
matrix structure all topological excitations of the string 
world-sheet.
The one-loop approximation of the model suggests that the 
space-time geometry dynamically undergoes a spontaneous 
dimensional reduction from ten to four dimensions - 
a spontaneous breaking of the ten dimensional
Lorentz symmetry. 
It was actually the first ever proposed candidate 
in string theory for a dynamical mechanism of dimensional reduction,
so the model has attracted the attention of 
many researchers. In the context of string theory it
is called IIB matrix model or IKKT model after the authors \cite{ikkt}.

The conjecture about the spontaneous symmetry breaking
has been derived in the one-loop approximation of the model 
\cite{aikktt, aikkt}.
This approximation leads in the low energy limit to an effective model 
of graphs which are dressed with vector fields representing 
positions of string world-sheet points
in space-time. The graphs are closely related to
branched polymers which have fractal dimension four 
\cite{adj, aikk, bepw}. The essence of the conjecture
is that starting with a ten dimensional model
one ends up with a four dimensional one \cite{aikktt, aikkt}.

It has become crucial to check whether 
the one-loop approximation is a good approximation 
in the limit of large distances and whether 
the scenario holds beyond the one-loop 
approximation \cite{ns,kkkms,vw,kkks}. 
In this paper we study the first issue.

The model is not solvable analytically and it is also
very complicated to treat numerically because of a sign problem 
which appears when one integrates out the fermionic degrees of 
freedom \cite{an}.
Many approximation techniques have been developed to investigate
the behaviour of the model \cite{aikktt,aikkt,st,os,s,aabhn2,bbpt}.
Because of the complexity of the ten dimensional model
its four and six dimensional counterparts 
have been studied
to gain deeper insight into the geometrical nature
of the problem and the limitations of effective
low energy models \cite{aabhn1,aabhn2,bbpt,bpt}. 
The four dimensional case is much easier
since it is free of the sign problem. In four dimensions
one has discovered some universal properties of the model 
which also hold in higher dimensions.
For instance one has found a relation between the existence of
elongated spiky world-sheet configurations and the 
singularities of the partition function.
A simple argument about entropy of spiky geometries 
derived in four dimensions could be extended
to higher dimensions. It has allowed for deducing the proper 
singularity type of the partition function, unravelling
the geometrical structure of configurations responsible 
for the singularities \cite{bbpt}.

In this paper we also concentrate on the four dimensional
case to analyse the geometrical behaviour
of the low energy effective model 
given by a model of branched polymers obtained from
the supersymmetric Yang-Mills matrix model 
in the one-loop approximation.
We numerically study the gyration radius, the two-point correlation 
function and the Polyakov-line operator whose effective 
form was derived analytically in \cite{bpw} . 

Four dimensional supersymmetric Yang-Mills integrals are interesting
per se and in connection with supersymmetric QCD. Actually the
relation of reduced integrals to the $1/N$-expansion was 
first discovered in QCD \cite{ek} where it was also 
applied to study the zero momentum content of the theory \cite{gk}.

\section{The model}
The partition function of the model is given by a reduced
supersymmetric Yang-Mills integral \cite{ikkt,fkkt,aikktt}
\eqn
\label{partfunct}
Z&=&\int DA\, D\bar\Psi\,D\Psi\ e^{-{S}[A,\bar\Psi,\Psi]}\quad,
\eqnx
with the action 
\eqn
\label{mmaction}
S[A,\bar\Psi,\Psi]&=&
-\frac{1}{4g^{2}}Tr\,[A^{\mu},A^{\nu}]^{2}
-\frac{1}{2g^{2}}Tr\,\bar\Psi^{a}\,\Gamma^{ab}_{\mu}[A^{\mu},\Psi^{b}]\quad,
\eqnx
where $A^{\mu}\,(\mu=1,\ldots D)$ are traceless $N\times N$ Hermitian matrices
and $\bar\Psi^{a},\Psi^{a}$ are $N\times N$ traceless matrices of
Grassmannian variables, transforming  
as Majorana-Weyl spinors for $D=10$,
or as Weyl spinors for $D=4$. We shall discuss here the $D=4$ case. 

One is interested in the large $N$ behaviour of the model.
It can be studied by approximate methods. 
In particular one can use perturbation theory 
to estimate the contribution of quantum fluctuations 
around classical solutions \cite{aikktt,aikkt}. 
The main idea consists in splitting the fields  
$A^{\mu},\,\bar\Psi^{a}$ and $\Psi^{a}$ into classical 
part and quantum fluctuations as
\eqn
A^{\mu}_{ij}=x^{\mu}_{i}\delta_{ij}+a^{\mu}_{ij}\
\quad&\bar\Psi^{a}_{ij}=\bar\xi^{a}_{i}\delta_{ij}+\bar\psi^{a}_{ij}&\ 
\quad \Psi^{a}_{ij}=\xi^{a}_{i}\delta_{ij}+\psi^{a}_{ij}\ ,
\eqnx   
and integrating out the quantum fluctuations $a$ and $\psi$.
In addition,
it is also convenient to integrate out
the classical fermionic fields $\xi$ to get rid of 
the Grassmannian content in the effective model.
Having done this one is led to a model of branched polymers 
with the partition function
\eqn
Z_{\mbox{\scriptsize{bp}}}&=&
\sum_{T\in{{T}}_{N}}\int\prod_{i}d^4x_{i}\ 
e^{-{S}_{\mbox{\scriptsize{bp}}}[T,x]}\quad,
\label{Z}
\eqnx
which describes the large-distance behaviour of the
semiclassical fields $x_i^\mu$ (we shall write in short 
$\vec{x}_i$) representing the positions 
of string world-sheet points in the target space.
If the positions are widely separated, 
that is $|\vec{x}_{i}-\vec{x}_{j}|\gg\sqrt{g}$, 
the effective action is given by \cite{aikktt,aikkt}
\eqn
S_{\mbox{\scriptsize{bp}}}[T,x]=
6\sum_{\langle ij\rangle\in T}\ln|\vec{x}_{i}-\vec{x}_j|\quad.
\label{act}
\eqnx
A detailed analysis of the matrix model shows that
a strong repulsion occurs whenever any
two vectors $\vec{x}_i$ and $\vec{x}_j$ come close to
each other \cite{aikktt,aikkt}. 
One believes that details describing the short-distance 
repulsion do not affect the universal large-distance properties
of the system. One can therefore choose the way in which
one models the repulsion as long as it is short-ranged.
One can for instance use a hard core repulsion, which is relatively
easy to implement. In this case the action (\ref{act}) reads
\eqn
\label{hcore}
S_{\mbox{\scriptsize bp}}^{c}[T,x]&=&\left\{
\begin{array}{lcl}
6\sum\limits_{\langle ij\rangle\in T}\ln|\vec{x}_{i}-\vec{x}_{j}
| & & \mbox{if}\quad |\vec{x}_{a}-\vec{x}_{b}|>c
\quad\forall\ a,b\in T\\
+\infty & & \mbox{otherwise}
\end{array}\right.
\eqnx
The size of the core $c$ is a free parameter in the model.
If any two vertices of the branched polymer 
come too close to each other the action becomes infinite
and the corresponding configuration is entirely suppressed
in the partition function. Each vertex is surrounded by
a ball-shaped zone which contains no other vertices of the 
branched polymer. From the numerical point of view
the branched polymer model is much easier to simulate than
the matrix model. The complexity
of the algorithm, defined as the number of operations
to update all degrees of freedom
grows as $N^8$ for the $SU(N)$ 
supersymmetric Yang-Mills model (\ref{partfunct})
and as $N^2$ for the branched polymer model (\ref{Z}). 
In the first case the number of degrees of freedom
which one has to update in one sweep is proportional
to the number of elements of the $A^\mu$ matrices, and
grows as $N^2$. In an update of an element of the matrix
one has to compute a determinant of the Dirac operator
which is a $K \times K$ matrix of size $K \propto N^2$.
Computation of a determinant requires 
$\propto K^3$ operations, yielding
altogether an at least $N^8$ complexity of a sweep in the matrix
model. In the second case in a sweep one has to update 
$N$ vertices. In each update one has to make 
$N-1$ global checks of the hard core condition
with remaining vertices. This yields an $N^2$ complexity
of a sweep in the branched polymer model. 
The presence of a hard core leads to an additional
slowing down of the algorithm:
Branched polymers have generically fractal
dimension four. Thus if they are embedded
in four dimensions they densely fill up the space.
Because of the hard core constraints the configuration space 
looks like a cheese with empty holes and the algorithm has 
to maneuver between them while 
constructing new configurations. Since the algorithm
does it by trial and error it takes a long time to 
produce completely uncorrelated configurations. Even
if one takes the autocorrelations into account 
the complexity of the branched polymer algorithm is many
orders of magnitude smaller than of the corresponding
matrix model algorithm for the same $N$,
already when $N$ is of order ten.

\section{Physical quantities}

Physical quantities 
are defined as operators $O[A^\mu,\Psi]$
which depend on the matrices $A^\mu$ and $\Psi$
in the original model (\ref{partfunct}).
In the one-loop approximation, after integrating
out quantum fluctuations $a^\mu$, $\psi$
and classical fermionic fields $\xi$, the
effective operators $O[A^\mu,\Psi] \rightarrow O_{bp}[x]$
become functions of a branched polymer
graph $T$ and fields $\vec{x}_i$ dressing its
vertices, which we shall denote in short by $x$. 
For each operator one has to derive its branched polymer counterpart.
The Polyakov-line operator, which is the fundamental operator
in the theory
\eqn
P_k = \frac{1}{N} \mbox{Tr} \exp \left[ i k_\mu A^\mu \right]
\eqnx
is mapped into the following operator in the branched 
polymer picture \cite{bpw}:
\eqn
\label{pl}
{{P}}_{bp,k}[x]&=&\frac{1}{N}\Bigg(\sum_{a}e^{ik_{\mu}x^{\mu}_{a}}
-k^{2}\sum_{a<b} 
\frac{e^{ik_{\mu}x_{a}^{\mu}}-e^{ik_{\nu}x_{b}^{\nu}}}
{ik_{\rho}(x_{a}^{\rho}-x_{b}^{\rho})|x_{a}-x_{b}|^{2}}\\[0.3cm]
&-& \sum_{\langle ab\rangle\in T}
\frac{e^{ik_{\mu}x_{a}^{\mu}}-e^{ik_{\nu}x_{b}^{\nu}}}
{ik_{\rho}(x_{a}^{\rho}-x_{b}^{\rho})|x_{a}-x_{b}|^{2}}\cdot
\frac{2|x_{a}-x_{b}|^{2}k^{2}-((x_{a}^{\sigma}-x_{b}^{\sigma})\,
  k_{\sigma})^{2}}{3|x_{a}-x_{b}|^{2}}\Bigg)\ . \nonumber
\eqnx
The operator contains three terms. The first one has a classical
origin. It is given by a sum over branched polymer vertices.
The second contains a sum over all pairs of vertices 
independently of whether they are neighbours on the branched 
polymer or not, while the third one - a sum over pairs 
of neighbouring vertices on the branched polymer graph.
These two terms come from quantum fluctuations 
in the one-loop approximation.

The most fundamental physical observable describing 
the distribution of the classical fields $\vec{x}_i$ 
in the target space is the two-point correlation function:
\eqn
G^{(2)}_N(\vec{X}_1,\vec{X}_2) = 
\left\langle \frac{2}{N(N-1)} 
\sum_{i<j} \delta(\vec{x}_i - \vec{X}_1) 
\sum_j \delta(\vec{x}_j - \vec{X}_2) \right\rangle_N \quad .
\eqnx
The average is taken over the ensemble $\{T,\vec{x}_i\}$
of branched polymers with $N$ vertices with the partition
function (\ref{Z}).
The statistical weight of the configuration is given by
a product of link weights 
\eqn
e^{-S} = \prod_{\langle ij \rangle \in T} W(|\vec{x}_i - \vec{x}_j|)
\eqnx
on which additionally the hard core constraints are imposed.
Since the statistical weight and the hard core constraints
are invariant with respect to a shift 
of all coordinates by the same vector 
$\vec{x}_i \rightarrow \vec{x}_i + \vec{\delta}$,
the two-point function $G^{(2)}_N$ depends on the difference
$\vec{X} = \vec{X}_2 - \vec{X}_1$, or actually - due to the isotropy
of the statistical weight - only on the distance $X = |\vec{X}|$.

Using the two-point function one can determine the typical linear
extent of the system. Usually one does it by calculating
the second moment of the two-point function:
\eqn
\langle X^2 \rangle_N = \int_0^\infty dX X^2 G^{(2)}_N(X)
= \left\langle \frac{2}{N(N-1)} \sum_{i<j} |\vec{x}_i - \vec{x}_j|^2
\right\rangle_N \ .
\label{x2}
\eqnx
The square root of this expression gives a quantity
called radius of gyration which is a standard 
measure of the linear dimension. For branched polymers
one can change integration variables in the partition
function (\ref{Z}) from $x$'s to $r$'s:  
$\vec{r}_{ij} = \vec{x}_i - \vec{x}_j$. Now one can see
that the integration over $\vec{r}_{ij}$'s in (\ref{x2})
leads to a divergence when it is done for
link weights with power-law tails 
$W(r) \sim r^{-6}$: the integration of $r^2$ over the four
dimensional volume $d^4 r = \Omega\, r^3 dr$, where $\Omega$
is the angular part of the integration measure,
gives a logarithmically divergent quantity:
$\langle r^2 \rangle = \Omega \int d r r^3 r^2\, W(r) 
\sim \int dr/r$. The gyration radius (\ref{x2}) is ill defined.
In this case the linear extent can be defined by 
the first moment of the two-point correlation function \cite{bepw,aabhn2}:
\eqn
R_N=\langle X \rangle_N = \int_0^\infty dX X G^{(2)}_N(X)
= \left\langle \frac{2}{N(N-1)} 
\sum_{i<j} |\vec{x}_i - \vec{x}_j| \right\rangle_N \quad .
\label{RN}
\eqnx
We will use this quantity in this paper to measure 
the linear extent of the system.

\section{Geometry of branched polymers}

We generate our configurations using a  Monte Carlo
algorithm. It consists of two update types which are applied
alternately:
a graph structure update
and a vertex positions update \cite{bepw}. 
When the branched polymer topology is updated vertex 
positions are kept constant, while when the vertex 
positions are updated the graph's topology
is kept constant.  
We tested the algorithm for Gaussian branched polymers with
link weights $W(r) = e^{-r^2}$, $r=|\vec{x}_i-\vec{x}_j|$
and without a core.
In this case one can determine the form of the two-point
function analytically \cite{bepw}. For large $N$
one expects the two-point function $G^{(2)}_N(X)$ to
effectively be a function of the scaling variable $X/N^{1/4}$.
In other words one expects that the fractal dimension is $D_f=4$.
One can see in figure \ref{tpf_gauss} that indeed the data
for the two-point functions for different $N$ collapse 
to one curve if one rescales $X \rightarrow X/N^{1/4}$.
\FIGURE{\epsfig{file=tpf_gauss+exact.eps,width=10cm}
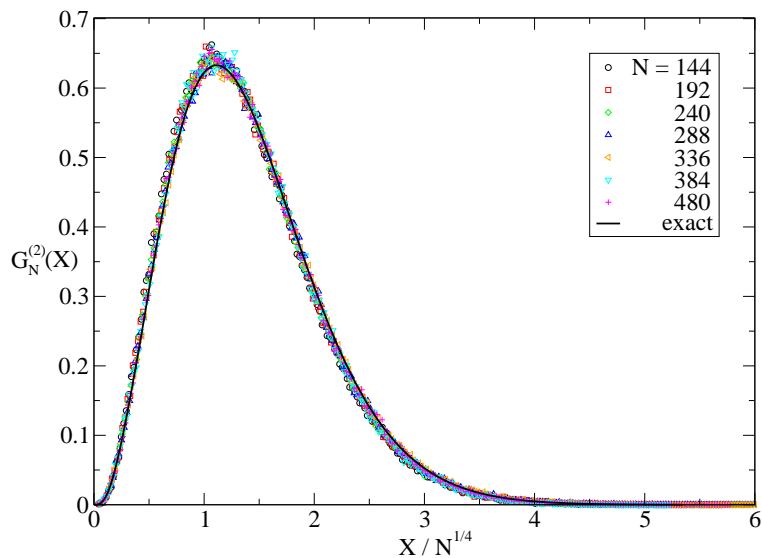
\caption{The two-point correlation function
$G_{N}^{(2)}(X)$ for the Gaussian branched polymer 
model plotted against $X/N^{1/4}$.\label{tpf_gauss}}
}
We now turn to the branched polymers (\ref{hcore})
derived from the matrix model (\ref{partfunct}).
The two-point function for different 
system sizes is plotted against the scaling variable in figure \ref{tpfc1}.
One can see small deviations from the scaling. 
\FIGURE{\epsfig{file=tpf_c1a0.eps,width=10cm}
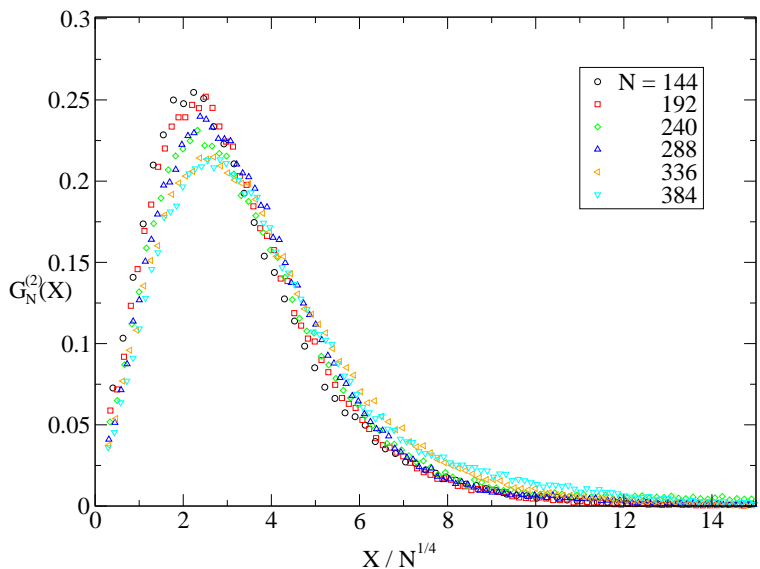
\caption{\label{tpfc1} The two-point correlation function 
$G_{N}^{(2)}(X)$ for the branched polymer model 
with power-tail weights and a hard core $c=1$ 
plotted against $X/N^{1/4}$.}}
The reason for this behaviour is related to the presence 
of the power-law tail in the link length distribution $W(r)$
which leads to the divergence of
the second moment of the distribution 
$\langle r^2 \rangle = \int d^4 r \, r^2 W(r) \sim \int dr/r$ 
as already mentioned above. 
This means that fluctuations of the link lengths are infinite 
and that from time to time long links appear in the system. These
links are reminiscent of the spiky configurations observed in
the surface model \cite{bbpt}. Let us now try to understand algorithmical
problems caused by the presence of such a fat-tailed distribution.
The algorithm which we use is an example of dynamical Monte Carlo.
It generates a Markov chain of configurations
randomly walking in the configuration space. Any two
consecutive configurations in the chain differ from each other
only a little.
In a single step the algorithm changes a link vector by a small 
value $\vec{r} \rightarrow \vec{r} + \vec{\delta}$ 
and accepts
this change with the Metropolis probability. The sequence of 
changes can be viewed as a sort of random walk in the
potential $\ln W(r)$. This potential is very flat 
and therefore if the algorithm once produces 
a long link it takes a long time to make it short
again. Such a random walk algorithm introduces 
therefore large autocorrelation times.
Moreover, a long excursion towards the tail of the
distribution produces many long links, so after each long excursion
a surplus of long links and a deficit of short ones 
occurs in the recorded history. This is an effect which makes
the rescaled histograms in figure \ref{tpfc1} not to
coincide. The narrow tails of the histogram
go far beyond the range in the figure. The tail behaviour
of the two-point function
is presented in the logarithmic scale in figure \ref{mmobpc1}.
It is compared with the tail
behaviour of the original matrix model 
(\ref{partfunct}). The two-point correlation function
inherits the power-law tail $dr r^{-3}$ from the link
length distribution $\sim dr r^3 W(r)$. As we can see in 
figure \ref{mmobpc1} in both cases the data behave as $\sim r^{-3}$.
\FIGURE{\epsfig{file=comp_mmown_bp_c1.eps,width=10cm}
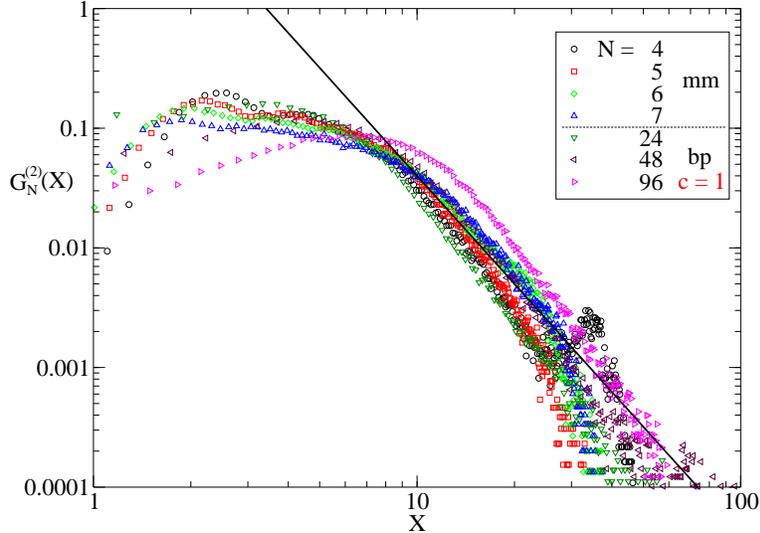
\caption{\label{mmobpc1}Comparison of the two-point function
$G_{N}^{(2)}(X)$ of the branched polymer model with power-tail weights
regularised by introducing a hard core $c=1$ and the IIB matrix model for
different values of $N$. The solid black line illustrates 
the expected power-law behaviour: $\sim X^{-3}$.}}

The presence of the tails $\sim r^{-3}$ makes the determination of
the linear extent of the system $R_N$ (\ref{RN}) difficult.
The computation of $R_N$ amounts to the computation of
the first moment of the two-point function.
The autocorrelation time 
for the algorithm is large and grows with $N$. 
On the other hand one expects
using the central limit theorem that the tail
$r^{-3}$ belongs to the Gaussian universality class,
what means that the probability of entering the
tail part of the distribution decreases with the
number of degrees of freedom, in our case with $N$.
One therefore expects that in the limit $N\rightarrow \infty$ 
the quantity $R_N$ should depend on the bulk of the distribution 
and not on the tail which becomes marginal in this limit,
and thus that the broadening of the statistical
error coming from the large autocorrelation time 
should be finite for $R_N$ for large system sizes.
One can use standard methods to estimate the error bars 
of $R_N$. If we do so we obtain the data presented 
in figure \ref{mrnewc1} illustrating the dependence 
of the linear extent $R_N$ on the system size. 
\FIGURE{\epsfig{file=meanrnew.eps,width=10cm}

\caption{\label{mrnewc1}
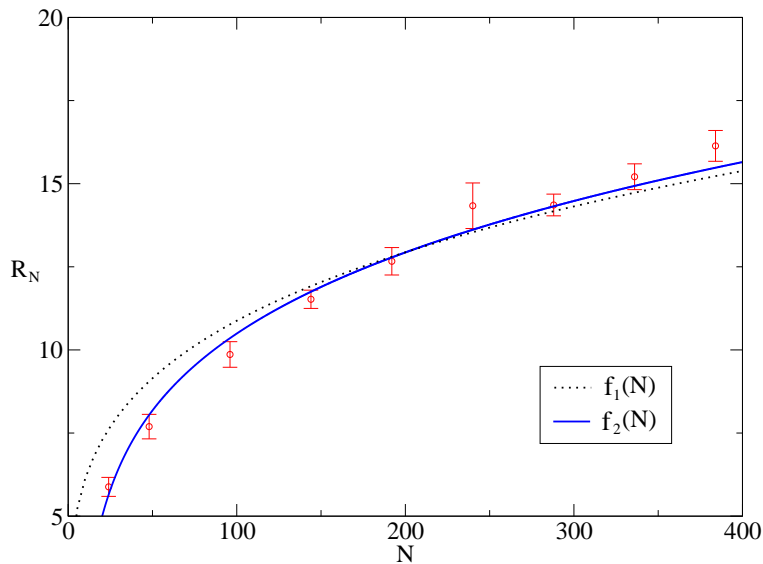
Comparison of the average gyration radius $R_{N}$ 
for the branched polymer model with a hard core $c=1$ plotted against 
the number of vertices $N$. The continuous lines represent the
functions:
$f_{1}(N)=3.44\cdot N^{1/4}$ and 
$f_{2}(N)=3.56\cdot N^{1/4}(1-6.8N^{-1})$.}}
The data presented in figure \ref{mrnewc1} can be fitted with the
scaling formula $f_{1}(N)=a_{1}\cdot N^{1/4}$.
If we apply this formula to find the best fit for the
range $N>48$ we obtain $a_{1}=3.44(5)$ with 
the fit quality: $\chi^{2}_{dof}=2.9$. 
If we include finite size corrections 
$f_{2}(N)=a_2 \cdot N^{1/4}(1+b_2 N^{-1})$, then the best fit
is for $a_{2}=3.56(4)$ and $b_{2}=-6.8(9)$ and has the
quality $\chi^{2}_{dof}=1.1$. The fit quality improves in
comparison with the previous one as can also be seen 
with the bare eye. The finite size corrections which control
deviations from the pure scaling can be attributed to 
the power-law tails which now and then cause the appearance 
of a long link on the branched polymer
and to the hard core repulsion which also
effectively makes the polymer look more elongated. 
The repulsion disfavors crumpled trees.
In principle one could expect the fractal dimension 
of branched polymers to lower.
The magnitude of the finite size corrections in
$f_2(N)$ is for $b_2 = -6.8$ of order a few percent for 
$N$ is of order a few hundred. 
One can thus believe that the results are consistent 
with the fractal dimension $D_f=4$.
To thoroughly check this hypothesis one would though 
have to study much larger systems. 

\section{Polyakov-line operator in the branched polymer model}

In this section we will discuss 
measurements of the Polyakov-line operator.
As it stands in equation (\ref{pl}) it has
real and imaginary parts. The imaginary part however
gives zero on average due to the reflection symmetry:
the partition function of the branched polymer model is
invariant with respect to simultanous reflections
$\vec{x}_{i}\rightarrow -\vec{x}_{i}$ 
of all vertex positions and therefore the average value of 
the Polyakov-line operator over the ensemble 
of branched polymers fulfills the condition:
\eqn
\left\langle P_{bp,k}[x] \right\rangle=
\left\langle P_{bp,k}[-x] \right\rangle .
\eqnx
Thus it is sufficient to measure only 
the real part of the operator. 
We shall denote it by $P_{k}$. The momentum $\vec{k}$ plays 
the role of an external parameter. Since the branched polymer system 
is isotropic one can choose $\vec{k}$ to lie on one of the axes 
of the coordinate system in which components of
the fields $\vec{x}_i$ are expressed.
In practice, while doing numerical simulations one can
calculate projections on the four independent 
components $\mu=1\ldots4$ and average over them 
to improve statistics.
This procedure eventually leads to the following operator
\eqn
\label{emm}
{P}_{k}[x]&=& P^{(1)}_{k}[x] + P^{(2)}_k[x] + P^{(3)}_k[x] 
\eqnx
where
\eqn
P^{(1)}_{k}[x] &=&
\frac{1}{4N}\sum_{\mu=1}^{4} 
\sum_{a}\cos(kx^{\mu}_{a}) \ ,
\nonumber \\
P^{(2)}_{k}[x] &=&
\frac{-k}{4N}\sum_{\mu=1}^{4} 
\sum_{a<b} \frac{\sin(kx_{a}^{\mu})-\sin(kx_{b}^{\mu})}
{(x_{a}^{\mu}-x_{b}^{\mu})|x_{a}-x_{b}|^{2}} \ , \nonumber\\
P^{(2)}_{k}[x] &=&
\frac{-k}{4N}\sum_{\mu=1}^{4} 
\sum_{\langle ab\rangle\in T}
\frac{\sin(kx_{a}^{\mu})-\sin(kx_{b}^{\mu})}
{(x_{a}^{\mu}-x_{b}^{\mu})|x_{a}-x_{b}|^{2}}\cdot
\frac{2|x_{a}-x_{b}|^{2}-\left( x_{a}^{\mu}-x_{b}^{\mu}\right)^{2}}
{3|x_{a}-x_{b}|^{2}} \ .\nonumber
\eqnx
In the last equations $k$ stands for the length of the
vector $\vec{k}$, which is independent of the space-time
direction $\mu$. In particular $k x_\mu$ is a product
of the length of the vector $\vec{k}$ 
and the $\mu$-th component of $\vec{x}$.

The length scale in the branched polymer model is set by the
size of the core $c$. It is the only scale parameter in
the model. The model is invariant under
a simultaneous rescaling $c \rightarrow \lambda c$
and $\vec{x}_i \rightarrow \lambda \vec{x}_i$.
One can use this invariance to rescale the core size
to $c=1$. The average of an operator $O[x]$
over the ensemble of branched polymers with a hard core 
of size $c$ can be expressed as the average of this
operator for the rescaled argument $O[cx]$ over the system
with a core $c=1$:
\eqn
\langle O[x] \rangle_c = \langle O[cx] \rangle_{c=1}  \ .
\eqnx
In particular for an operator which 
is a homogeneous function of order $\Delta$: 
$O[\lambda x] = \lambda^{\Delta} O [x]$
we have
\eqn
\langle O[x] \rangle_c = c^{\Delta} \langle O[x] \rangle_{c=1} \ .
\eqnx
The momentum $k$ of the Polyakov-line operator introduces an additional scale
which has to be measured relatively to $c^{-1}$.
If one rescales $\vec{x} \rightarrow \lambda \vec{x}$
one should simultaneously rescale $k \rightarrow k/\lambda$ to keep
the combination $\vec{k} \vec{x}$ constant. 
We can now relate the value of the Polyakov-line in the
system with a hard core $c$ to its value in the system with $c=1$:
\eqn
\label{ppp}
\left\langle P^{(1)}_{k}[x] \right\rangle_c = 
\left\langle P^{(1)}_{ck}[x] \right\rangle_{c=1} \ , \quad
\left\langle P^{(2,3)}_{k}[x] \right\rangle_c = 
c^{-4} \left\langle P^{(2,3)}_{ck}[x] \right\rangle_{c=1} \ .
\eqnx
The amplitudes of the second $P^{(2)}_k[x]$ and 
the third term $P^{(3)}_k[x]$ scale 
as $c^{-4}$ with  the core size 
while the amplitude of the first term $P^{(1)}_k[x]$
stays constant. Using this observation we can
reconstruct the behaviour of the average of the
Polyakov-line operator $\langle P_k\rangle_c$
in the system with an arbitrary $c$ from measurements of 
$\langle P^{(1)}_k\rangle$, 
$\langle P^{(2)}_k \rangle$ 
and $\langle P^{(3)}_k\rangle$ for $c=1$. 
We will follow this strategy. In figure \ref{poly123}
we present the dependence of
the three contributions for $c=1$ on the momentum $k$,
obtained by numerical Monte Carlo measurements.
\FIGURE{\epsfig{file=poly123.eps,width=10cm}
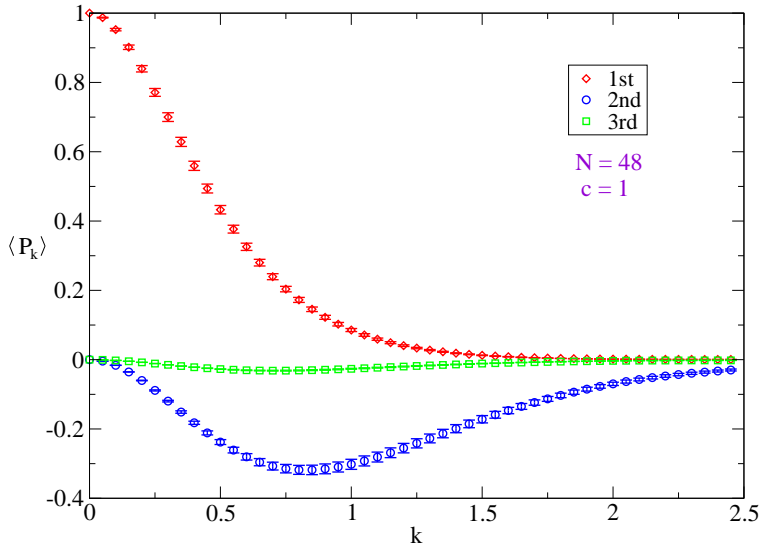
\caption{The three terms $\langle P^{(1)}_k\rangle_{c=1}$, 
$\langle P^{(2)}_k \rangle_{c=1}$ 
and $\langle P^{(3)}_k\rangle_{c=1}$ contributing to
the average Polyakov-line operator in the
one-loop approximation (\ref{emm}) in the system with hard core size $c=1$. \label{poly123}}}
Since the second and the third terms are negative,
the Polyakov-line operator being 
a linear combination of the three terms 
(\ref{emm}) may assume negative values for some
range of $k$ depending on the value of $c$ (\ref{ppp}). 
Such a behaviour is also observed in the matrix model
for small $N$ as we can see in figure \ref{su2po}. 
In the matrix model the effect however disappears 
when $N$ increases and is practically
absent already for $N$ of order ten. 
The curve for $N=2$ in figure \ref{su2po}
was obtained analytically as a Fourier
transform of the eigenvalue distribution whose form 
is known for $SU(2)$ \cite{kps}. 
\FIGURE{\epsfig{file=test_mm2_thin.eps,width=10cm}
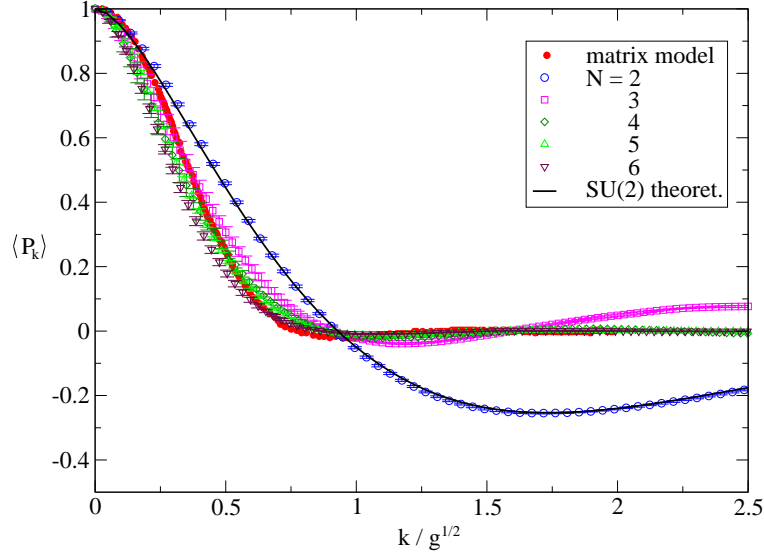
\caption{\label{su2po}Comparing the Polyakov one-point 
correlation functions $\langle P_{k}\rangle$ of the 
IIB matrix model for $N=2,\,3,\,4,\,5,\ 6$.}}

Now the issue is to test up to what degree the model 
of branched polymers can reproduce the full matrix model. 
The model of branched polymers was constructed
as an effective model for the large-distance behaviour
of the original model. So we are now looking for
the optimal value of the parameter $c$, which is 
a free parameter of the effective model, 
to reproduce the behaviour of the matrix model on large 
distances which correspond to small values of $k$ in
the Polyakov-line operator.
For the comparison we will use the matrix model results
presented in \cite{aabhn2}. They correspond to
$N=16,\,24,\,32$ and $48$.  The coupling constant $g$ of the matrix
model is chosen to be $g=\sqrt{48/N}$ and the results are expressed in
physical units $k_{\mbox{\tiny{phys}}}=k/\sqrt{g}$ \cite{aabhn2}. 
After rescaling to physical units the data of the
matrix model for different $N$ collapse to one curve.
This curve goes as $1- \alpha k^2 + \dots$ for small $k$.
We found that for $c=0.75$ also the branched polymer
results follow this curve in the region of small momenta.
The results are presented in figure \ref{ck2}.
\FIGURE{\epsfig{file=polyline_c075_k2.eps,width=10cm}
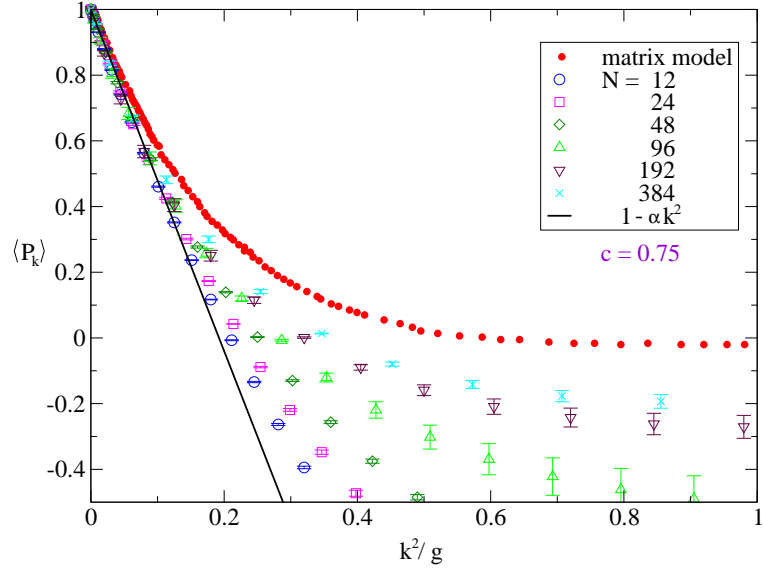
\caption{\label{ck2}Comparing the $1 - \alpha k^2 + \dots$
behaviour of the Polyakov-line operator
$\langle P_{k}\rangle$ of a branched polymer
with $c=0.75$ for different sizes of $N$ 
and the IIB matrix model results.}} Indeed for small $k$ the data points for different $N$
follow the same line $1- \alpha k^2$.
The slope parameter $\alpha$ is thus universal. For larger
$k^2$ the results depart from the universal line in a non-universal
manner which depends on $N$. There are two
reasons for that. First of all the effective model 
by construction is only supposed to describe the
large-distance (small $k$) behaviour. 
For larger $k$ the effective model depends on 
the details of the regularization and is not any more
universal. Secondly the Polyakov-line operator (\ref{pl}) 
contains only one-loop corrections.
Higher loop corrections, which have been neglected in (\ref{pl})
would introduce terms of order $O(k^4)$. 
Let us stress again, what is most important in figure \ref{ck2}
is that the leading $k^2$ behaviour is universal
and identical with that of the original matrix model. 

\section{Summary}
We have shown that the branched polymer model very well
describes all essential features of the reduced
supersymmetric Yang-Mills integrals in four dimensions
at large distances: the power-law tail behaviour of the 
two-point function, the four dimensional 
scaling of the gyration radius and the universal
small momentum behaviour of the Polyakov-line operator.
The branched polymer model can thus be useful in studies
of the low momentum content of supersymmetric QCD \cite{gk}.

What remains to be checked is whether the one-loop approximation
works equally well for the ten dimensional matrix model where
branched polymers are replaced by somewhat more
complex graphs in the effective model. These graphs
have branched polymer backbones and one believes
that they preserve the four dimensional fractal structure.
Additional links which appear on the graphs
introduce some rigidity constraints which therefore may
make the fractal dimension to become the canonical dimension
of the underlying geometry. This would explain the existence
of a dynamical
spontaneous symmetry breaking of the original ten dimensional
Lorentz symmetry. There are some indications that such a scenario indeed
takes place \cite{ns,kkkms,vw,kkks,an,aabhn3,nos}.

\bigskip
\noindent
{\bf Acknowledgments}

\medskip
\noindent
This work was partially supported by the Deutsche Forschungsgemeinschaft 
under the project FOR 339/2-1, 
the Polish State Committee for Scientific Research (KBN) grant
2P03B-08225 (2003-2006) and by EU IST Center of Excellence ``COPIRA''.

\end{document}